\pdfoutput=1
\documentclass[acmtog,authorversion,preprint]{acmart}

\usepackage{booktabs}
\usepackage{epsfig}
\usepackage{graphicx}
\graphicspath{{./figures/}}
\usepackage[utf8]{inputenc}
\usepackage{siunitx}
\usepackage[T1]{fontenc}
\usepackage{ctable}
\usepackage{array}
\usepackage{anyfontsize}
\usepackage[inline]{enumitem}
\usepackage{comment}
\usepackage{amsmath}
\usepackage{amsfonts}
\usepackage{xr}
\usepackage{multirow}
\usepackage{color,soul}
\usepackage{bm}

\def\figurePath{figures/}

\def\myfigure#1#2{\begin{figure}[t]\centering\includegraphics*[width = \linewidth]{\figurePath#1}\caption{#2 }\label{fig:#1}\end{figure}}
\def\mycfigure#1#2{\begin{figure*}[t]\centering\includegraphics*[clip, width = \linewidth]{\figurePath#1}\caption{#2 }\label{fig:#1}\end{figure*}}

\def\mysection#1#2{\section{#1}\label{sec:#2}}

\def\mysubsection#1#2{\subsection{#1}\label{sec:#2}}

\newcommand{\refSec}[1]{Section~\ref{sec:#1}}
\newcommand{\refFig}[1]{Figure~\ref{fig:#1}}
\newcommand{\refEq}[1]{Equation~\ref{eq:#1}}
\newcommand{\refTbl}[1]{Table~\ref{tbl:#1}}

\newcommand{\degree}{^\circ}
\newcommand{\cpd}{\,\si{cpd}}
\newcommand{\Hz}{\,\si{Hz}}

\DeclareMathOperator*{\argmin}{arg\,min}

\newcommand{\irow}[1]{%
	\begin{smallmatrix}[ #1 ]^\top\end{smallmatrix}%
}

\usepackage[ruled]{algorithm2e}

\SetAlFnt{\small}
\SetAlCapFnt{\small}
\SetAlCapNameFnt{\small}
\SetAlCapHSkip{0pt}

\acmJournal{TOG}

\setcopyright{none}
\renewcommand\footnotetextcopyrightpermission[1]{} %
\makeatletter
\renewcommand\@formatdoi[1]{\ignorespaces}
\makeatother

\makeatletter
\def\runningfoot{\def\@runningfoot{}}
\def\firstfoot{\def\@firstfoot{}}
\makeatother

\begin{document}

\title{Perceptual Visibility Model for Temporal Contrast Changes in Periphery}

\author{Cara Tursun}
\email{cara.tursun@rug.nl}
\affiliation{%
  \institution{Università della Svizzera italiana}
  \country{Switzerland}
}
\affiliation{%
  \institution{University of Groningen}
  \country{Netherlands}
}
\author{Piotr Didyk}
\email{piotr.didyk@usi.ch}
\affiliation{%
  \institution{Università della Svizzera italiana}
  \country{Switzerland}
}

\begin{teaserfigure}
\includegraphics[width=\textwidth]{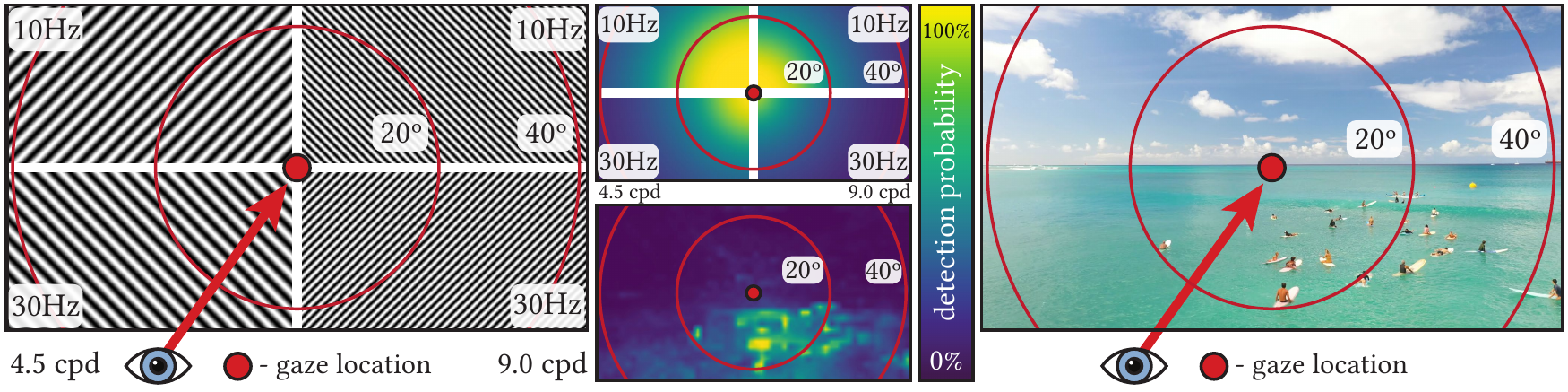}
\caption{Our technique predicts the visibility of temporal image changes across a wide field of view. The model takes into account both spatial and temporal frequencies of the content as well as the eccentricity. On the left, we show the temporal fluctuations of sinusoidal patterns with spatial frequencies of 4.5 and 9 cycles per visual degree across $40^\circ$ field-of-view for two different temporal frequencies: 10 and 30\si{Hz}. In the middle-top, we show our prediction for each quadrant of this input. The spatial frequencies were downscaled by a factor of 10, and the contrast was enhanced for better visibility. Our visibility prediction is shown in the middle-bottom for a natural video with surfers that is shown on the right.}
\end{teaserfigure}

\begin{abstract}
Modeling perception is critical for many applications and developments in computer graphics to optimize and evaluate content generation techniques. Most of the work to date has focused on central (foveal) vision. However, this is insufficient for novel wide-field-of-view display devices, such as virtual and augmented reality headsets. Furthermore, the perceptual models proposed for the fovea do not readily extend to the off-center, peripheral visual field, where human perception is drastically different. In this paper, we focus on modeling the temporal aspect of visual perception in the periphery. We present new psychophysical experiments that measure the sensitivity of human observers to different spatio-temporal stimuli across a wide field of view. We use the collected data to build a perceptual model for the visibility of temporal changes at different eccentricities in complex video content. Finally, we discuss, demonstrate, and evaluate several problems that can be addressed using our technique. First, we show how our model enables injecting new content into the periphery without distracting the viewer, and we discuss the link between the model and human attention. Second, we demonstrate how foveated rendering methods can be evaluated and optimized to limit the visibility of temporal aliasing.
\end{abstract}


\maketitle

\mysection{Introduction}{introduction}
The continuous improvements in display technology increase our ability to meet the perceptual capabilities of human visual perception, leading to a more realistic, engaging, and immersive user experience. Unfortunately, hardware developments also lead to more challenges regarding content creation and optimization techniques, which have to resolve multiple different quality trade-offs. At the forefront of these efforts are perceptually inspired techniques, which, informed by studies of human perception, aim at providing optimal user experience given hardware and computational limitations of current display systems. The success of such techniques has been demonstrated for many problems in computer graphics \cite{masia2013,weier2017}.

Many perceptually inspired techniques are based on variation in human sensitivity to spatio-temporal luminance patterns throughout the visual field \cite{barten1993}. In the past, consideration of spatial properties of the human visual system (HVS) led to many developments in image enhancement and rendering, for example, \cite{Krawczyk2007,Ramasubramanian1999}. Adding the temporal aspect allows handling complex phenomena governing the spatio-temporal aspect of the human perception and exploiting the human insensitivity to high temporal frequencies \cite{Yee2001,Didyk2010a,Didyk2010b,Berthouzoz2012}. Many of these aspects are parts of image and video metrics \cite{Mantiuk2011,Aydin2010} which are important for evaluating computer graphics techniques \cite{Narain2015,Gharbi2016,Andersson2020} and guiding optimization techniques \cite{Oeztireli2015,Wolski2019}.

Until recently, such techniques solely focused on addressing the perception of central vision, the so-called fovea. However, this turns out to be insufficient, especially for the new wide-field-of-view virtual and augmented reality headsets, which present high-resolution images that span a significant portion of the human visual field. Although these new capabilities enable both immersive virtual reality applications and high-quality real-world augmentation, the insufficient understanding of the processes governing the perception in the periphery prevents achieving the highest quality given limited computational budget. Accurate modeling of human visual perception in the periphery becomes even more critical for new display systems equipped with eye-tracking technology that enables precise information about the gaze location. Such information opens new opportunities for optimizing image content locally according to the position of the image content in the visual field. In computer graphics, the most significant applications leveraging these capabilities are gaze-contingent rendering techniques \cite{guenter2012,patney2016,stengel2016,tursun2019,Murphy2001}, which exploit the decline in human visual sensitivity to distortions with increasing eccentricity. Unfortunately, the perceptual models used in these applications are usually limited to static content, and only very few consider the temporal properties of the HVS in the periphery \cite{Bailey2009,sun2018}. Consequently, the lack of techniques for modeling the human sensitivity to spatio-temporal signals across a wide field of view keeps us from using the full potential of the new type of devices.

In this work, we specifically study the sensitivity of the HVS to spatio-temporal luminance patterns in the periphery. To this end, we first describe a series of experiments conducted to measure the visibility of spatio-temporal patterns. Based on these measurements, we build an efficient model that predicts the visibility for complex luminance patterns. The method relies on discrete cosine transform (DCT), which is commonly used in video processing application and ease the adoption of our model to a large range of applications. Our experiments are tailored to this decomposition and contain DCT basis functions. Despite using simple patterns in the experiments, we demonstrate that by drawing inspirations from the mechanism governing human perception and previous literature from visual science, our model provides a good prediction for complex stimuli. 

We also show several opportunities and applications which our model enables. These include analyzing video sequences for detecting visible temporal changes, invisible injection of new content in the periphery that does not create a distraction for an observer, and evaluating and optimizing foveated rendering to prevent visibility of temporal aliasing. We also demonstrate a possible link between the prediction of our model and human attention. To summarize, the main contributions of this paper are:\vspace{-1ex}
\begin{itemize}
    \item perceptual experiments investigating the visibility of spatio-temporal patterns in the periphery,
    \item computational model based on DCT decomposition that predicts the visibility of temporal changes for complex video and animation content across a wide field of view, and
    \item applications of the model for creating and optimizing content for wide-field of view displays, including a new technique for subtle introduction of new content in periphery.
\end{itemize}

\mysection{Related Work}{related_work}
Below, we describe related studies and models of the visibility of spatio-temporal contrast patterns. We also summarize existing quality and visibility metrics, which extend these models to complex stimuli, and applications that utilize both the perceptual models and metrics.  

\paragraph{Spatio-temporal contrast}
Studying and modeling the perception of spatio-temporal contrast has received a lot of attention in both visual science and computer graphics. One of the earliest studies that considered the sensitivity of the HVS to temporally changing patterns is conducted by De Lange \shortcite{de_lange1952}, which provided the threshold modulation for a relatively small stimulus size of $2^{\circ}$. These initial measurements showed that the HVS has the peak temporal sensitivity around 10 \si{Hz}, with a sharper falloff towards higher temporal frequencies with a cutoff around 60 \si{Hz}. The sensitivity to low temporal frequencies were measured by Thomas and Kendall \shortcite{thomas1962}, which were obtained in natural viewing conditions in a room where the room lighting was modulated. These measurements contrasted the study of De Lange in terms of the stimulus size and they reported much lower sensitivities. These differences are later studied by Kelly \shortcite{kelly1964}, who identified different effects from stimulus size and time-average luminance level ($L_0$) of the stimulus for low ($< 10 \si{Hz}$) and high ($> 20 \si{Hz}$) frequency counterphase sine waves. As for stimulus size, they observed an inverse relation between the size ($> 2^{\circ}$) and temporal modulation sensitivity for low frequencies, whereas high-frequency sensitivity was relatively unaffected. On the other hand, high-frequency sensitivity was reduced by decreasing $L_0$, whereas it had little effect on low-frequency modulation sensitivity. The studies of temporal modulation sensitivity are followed by the derivation of the so-called spatio-temporal contrast sensitivity function by Robson \shortcite{robson1966}. An extensive survey of the perceptual studies on the perception of temporal stimulus and a model of temporal sensitivity is provided by Watson \shortcite{watson1986}. The model introduced in that study is characterized by a linear filter, probability summation over time and thresholds for temporal changes of brightness. Another spatio-temporal model of contrast sensitivity is proposed by Barten \shortcite{barten1993} based on the temporal behavior of the photoreceptor cells in the eye. More recently, Watson and Ahumada \shortcite{watson2016} introduced a spatio-temporal visibility model called the pyramid of visibility, which is based on the observation that the contrast sensitivity of the human eye being a linear function of spatial and temporal frequencies. This led to a simple linear parameterization of the visibility thresholds in this high-dimensional space for high temporal and spatial frequencies.

\paragraph{The critical flickering frequency}
The HVS retains the visual impression of a stimulus for a brief amount of time after the stimulus disappears. This perceptual phenomenon is due to low-pass filtering effects of the HVS
and it results in intermittent light above a temporal frequency threshold, called critical flickering frequency (CFF) being perceived as continuous. CFF increases linearly with $\log$-stimulus area \cite{granit1930} and $\log$-luminance \cite{ferry1892, porter1902, makela1994}) up to a saturation point and then remains constant. It also increases with retinal eccentricity up to $30\degree$--$60\degree$ followed by a fall off at the far periphery \cite{hartmann1979, montvilo1981, tyler1987}. CFF is usually measured for stimuli without spatial structure. However, in the recent work of Krajancich et al. \shortcite{Krajancich2021}, they provide CFF measurements in peripheral vision for spatial frequencies up to $ 2\cpd $. Our work considers spatial luminance frequencies up to approximately $9\cpd$. In addition, we provide a model tailored directly for the spatio-temporal signal decomposition (DCT) of complex videos.
In other work, Mantiuk~et~al.~\shortcite{Mantiuk2021} also address peripheral vision. Their work proposes a quality metric for a wide field of view video sequences. While similar in applications, our work focuses on the local visibility of temporal changes and not the overall quality of the content. Our work also provides direct measurements of the human sensitivity to well-defined and localized patterns whereas their model is trained on a video dataset. Moreover, their method computes visible quality differences with respect to a reference input, while our work does not require a reference for detecting visible temporal changes. We provide a more in-depth discussion and comparison to the works of Krajancich~et~al. and Mantiuk~et~al. in Section~\ref{sec:Discussion}.

\mysection{Overview}{overview}

The ability to detect temporal changes in a visual stimulus depends on several factors:

\begin{enumerate}
	\item Amplitude of temporal modulation of light
	\item Spatial frequency content of the stimulus
	\item Retinal position of the stimulus and the distance to the central vision (fovea)
	\item Wavelength of the light
	\item Average illumination intensity 
	\item Local adaptation to temporal changes
	\item Stimulus area
	\item Age and fatigue level of the observer
	\item Visual masking
    \item Eye movements
\end{enumerate}

We focus on modeling the prominent effects of (1), (2), and (3) for designing a perceptual model that computes the probability of detecting the temporal changes by a human observer. Our model works on luminance contrast computed from the visual stimulus using the colorspace of display (e.g., sRGB). The luminance conversion takes into account the wavelength of the light (4). However, we do not consider the visibility of isoluminant chromatic contrast patterns. As for illumination intensity, our model is calibrated for photopic viewing conditions. To avoid local adaptation effects \cite{ginsburg1966}, we used an experiment design where the duration of the observation time does not affect the responses (as opposed to procedures like the adjustment method). It is known that the number of cycles has an influence on the measured contrast threshold for spatial sine wave gratings \cite{hoekstra1974, howell1978, virsu1979, tyler2015}. In our experiments, we focus on the visibility of localized temporal changes and choose a constant stimulus size for all tested retinal eccentricities. Our model does not account for visual masking effects and the effects of eye movements on the contrast thresholds \cite{kelly1979,daly2001b,laird2006}.

In the next section, we provide the details of our psychophysical experiment procedure for measuring the spatio-temporal contrast sensitivity. In \refSec{model}, we introduce our model that is calibrated using our measurements. In \refSec{applications}, we show applications of our model to predicting temporal change detection, controlling the visibility of temporal image transitions, and benchmarking the visibility of temporal aliasing in foveated rendering. Our applications to image transition and temporal aliasing also serve as a validation of our model because we compare the visibility of stimuli predicted by our method with experimental data. In \refSec{Discussion}, we compare our work with relevant studies and conclude our paper.
\mysection{Experiments}{experiments}
To build a computational model predicting the visibility of temporal image changes in the periphery, we first collect perceptual data to which the model can be fitted. Since the model relies on DCT decomposition, we seek the information regarding the sensitivity of the HVS to different components of DCT decomposition (DCT basis functions) \cite{ahmed1974}, which in our spatio-temporal case, contain a different mixture of cross-modulated horizontal and vertical sinusoidal gratings.

\paragraph{Stimuli}
Each of the stimuli can be described using four parameters: horizontal spatial frequency $f_h$, vertical spatial frequency $f_v$, temporal frequency $f_t$, as well as eccentricity $e$. Sampling the four-dimensional space is required but challenging due to the time-consuming procedure of measuring visibility threshold for each of them. To acquire sufficient data while keeping the perceptual experiment feasible, we sampled each dimension at three locations. More precisely, we used 81 stimuli, each being a combination of $f_h \in \{0\cpd, 4.54\cpd, 9.06\cpd\}$, $f_v \in \{0\cpd, 4.54\cpd, 9.06\cpd\}$, $f_t \in \{20\Hz, 30\Hz, 60\Hz\}$, $e \in \{10\degree, 25\degree, 40\degree\}$. Unlike previous psychovisual studies, which measure the sensitivity to isolated sinusoidal gratings with different orientations, the combination of horizontal ($f_h$) and vertical ($f_v$) frequencies in our experiments results in cross-modulated patterns (\refFig{stimuli_v3}). The eccentricity values were chosen to minimize the risk of presenting the stimuli in the participant's blind spot, which is usually located around $15\degree$ eccentricity \cite{wandell1997}. Each of the stimuli was a 71\,\si{px} $\times$ 71\,\si{px} square containing a pattern windowed with a circle having $2\degree$ diameter and containing a small Gaussian falloff. The temporal modulation was temporal blending between each the pattern and its negative and it spanned  200\,\si{ms} which allowed to reproduce all the temporal frequencies on 120\,\si{Hz} display using 25 frames. The stimuli were defined in a linear luminance space. The temporal average of each stimuli was the background color corresponding 50\,\% of the max intensity of the display (83.165 cd/m$^2$).

Visual sensitivity increases as the envelope of a sinusiodal grating grows with an asymptote between 3--10 cycles of the underlying signal \cite{howell1978}. Majority of previous spatio-temporal models and datasets use a Gabor with an envelope that allows for 2--3 cycles. Different from these previous psychovisual studies, we opt to keep stimuli size fixed in our measurements because of our application based on the constant window size of DCT.

\mycfigure{stimuli_v3}{
Two example sets of the spatio-temporal stimuli used in our experiments. Each row corresponds to one stimuli, i.e., a sequence of images. The group on the left is an example of three stimuli with different spatial frequencies and low temporal frequency. The group on the right presents a similar selection of spatial frequencies for the highest temporal frequency considered.
}

\paragraph{Task}
Each task comprised of establishing the sensitivity of the observer in detecting one of the stimuli. To this end, the participants performed a threshold estimation task for each stimuli to find the minimal amplitude of the stimulus' pattern, e.g., contrast, which is visible. We used vPEST \cite{findlay1978} procedure with two-alternative-forced-choice (2AFC) pairwise comparison.
At each trial the participants was first asked to fixate at the target shown in the screen. Then, one spatio-temporal pattern was shown at a given eccentricity in the mid-height of the display, either on the left or right side from the screen center. For each trial the participants were asked to decide on which side, the pattern appears. Participants answered using arrow keyboard keys. Estimation of thresholds for all 162 stimuli was split into 12 sessions where the vPEST procedures were run in parallel, and at each step the current stimuli from a random procedure was shown. One session took approximately 20\,minutes, and the participants were asked to take a break whenever they experienced fatigue. The experiment protocol was approved by the ethical committee of the host institution.

\paragraph{Participants}
Four participants (20-40 years old) took part in these measurements. All had normal or corrected-to-normal foveal vision. Participants did not report any peripheral vision deficiencies (peripheral acuity is not tested separately). Two participants were the authors of the paper. Before the experiments, it was verified that none of the stimuli falls into the blind spot of the participants. This was tested separately for both eyes by showing a black circle in place of the stimuli.

\sloppy  %
\paragraph{Hardware}
The experiments were conducted using a gamma-corrected 55-inch LG OLED55CX, 120\si{Hz}, 4K display. The OLED technology provides sufficiently fast response time which was negligible in our experiments. For more details on the evaluation of the display, see the supplemental materials. The setup of the display was optimized to maintain constant peak brightness (167.33 cd/m$^2$) and contrast (494:1) over time. Participants carried out the experiment using a chin-rest 62 \si{cm} from the display in a room with the ambient light level at 700 \si{lx}.

\paragraph{Results}
\refFig{experiment_plots_v2} shows the thresholds estimated during the experiment averaged across all participants. It can be observed that the thresholds decrease for lower spatial and temporal frequencies. For large spatial and temporal frequencies, the estimated values because close to 0.5, which is the maximum contrast that can be represented on the display. In these cases, the threshold estimation procedure saturates as no larger contrast values can be considered. 

\mycfigure{experiment_plots_v2}{The average temporal contrast levels for 75\% detection measured in our psychovisual temporal change detection experiment (\refSec{experiments}). Each plot shows the thresholds with respect to the retinal eccentricity of the stimuli. The temporal frequencies are shown at the top of the plots and the sinusoidal spatial frequency content (pairs of horizontal and vertical frequencies in cpds) represented by each line is given in the legend.}

\mysection{Model}{model}

\begin{table}
	\centering
	\caption{Notation that we use in this paper}
	\label{tbl:symbols}
	\resizebox{\linewidth}{!}{%
	\begin{tabular}{cp{68mm}}
		\hline
		\textbf{Symbol} & \textbf{Description} \\
		\hline
		$f_t$ & Temporal frequency ($\log$-Hz) \\
		$f_h, f_v$ & Horizontal and vertical spatial frequencies ($\log$-cpd) \\
		$e$ & Eccentricity ($\log$-deg) \\
		$q(\cdot)$ & Quadratic function \\
		$S_{DL}(\cdot)$ & Polynomial fit to De Lange curve in $\log$-domain \\
		$S_{SP}(\cdot)$ & Zero-truncated De Lange curve using softplus function \\
		$S(\cdot)$ & Eccentricity-dependent spatio-temporal contrast sensitivity in $\log$ domain \\
		$T(\cdot)$ & The scaling function for the temporal contrast sensitivity curve \\
		$U(\cdot)$ & The function for shifting the temporal contrast sensitivity curve across the time axis ($f_t$)  \\
		$C(\cdot)$ & Band-limited spatio-temporal luminance contrast \\
		$C_{\text{JND}}(\cdot)$ & Just-Noticable Difference scaled spatio-temporal luminance contrast \\
		$C_M(\cdot)$ & JND-scaled contrast after spatial and temporal pooling \\
		$p_g$ & Guess rate, the probability of selecting the stimulus by pure chance in psychovisual experiment (e.g., by random guessing)\\
		$p_l$ & Lapse rate, the probability of not selecting the detected stimulus in psychovisual experiment (e.g., by human-error) \\
		\hline
	\end{tabular}%
	}
\end{table}

Our model is derived from the temporal contrast sensitivity function of De Lange \shortcite{de_lange1952}, which is measured for fovea. We start by representing the De Lange curve using a polynomial approximation in \refSec{spatiotemporal_sensitivity}. Then we present our DCT-based stimulus decomposition method in \refSec{stimuli_decomposition}. In \refSec{temporal_change_detection}, we describe the computation of change detection for the periphery and its calibration to the experimental data introduced in \refSec{experiments}. Finally, we calibrate our model using the data from our psychophysical experiment with complex stimuli consisting of video snippets from natural videos in \refSec{calibration}.

\mysubsection{Spatio-temporal sensitivity}{spatiotemporal_sensitivity}
In order to derive an observer's sensitivity to visual stimuli, we need the psychometric function that gives the subject's response to the different stimulus levels. In practice, it is possible to assume that the behavior changes smoothly around the measurement points from a psychovisual experiment and make a prediction from the experimental data. This approach is nonparametric, but it requires a large number of measurements in our case because of increased dimensionality of the psychometric function space, mainly due to the changes in spatial and temporal frequencies as well as the retinal eccentricity. Due to practical considerations for avoiding participant fatigue and keeping the length of experiment sessions short, it is not feasible to collect experimental data that fully span this high-dimensional space. Instead, we take an alternative approach and base our spatio-temporal sensitivity model on the De Lange curve, which represents the human visual system's sensitivity to temporal modulations of light at different frequencies \cite{de_lange1952}. Then we aim to model the effects of additional factors such as the retinal eccentricity and spatial frequency content using functions that control the shape of the curve. As a result, we have a more compact set of parameters that change the sensitivity in semantically meaningful ways such as by tuning the position of the peak sensitivity or how fast the sensitivity declines with the retinal eccentricity. This approach was also used in some of the previous psychovisual studies and it effectively reduces the amount of experimental measurements required to a plausible level \cite{lesmes2010, watson2016}.

We start by expressing the measurements of De Lange curve at fovea using a curve fit and then introduce an extension to the peripheral visual field and multiple spatial frequencies (please see \refTbl{symbols} for the notation that we use in this paper).

\paragraph{De Lange fit}
The curve fit is represented by a polynomial of degree $n$ in the $\log$-sensitivity and $\log$-frequency domain:
\begin{equation}
\label{eq:polynomial}
    S_{DL} (f_t) = \sum_{i=0}^{n} a_i \cdot ( f_t ) ^ i,
\end{equation}
where $S_{DL}$ is the $\log$-sensitivity (1/threshold) to the temporal modulations at $\log$-frequency $f_t$ and $a_i$ is the coefficient. The mathematical singularity observed while computing $\log(f_t)$ at $f_t = 0$ is handled by using a power transformation, which is a more general form of the standard $\log$-transformation as defined in Appendix \ref{appendix:power_transform}.

The polynomial fit as given in \refEq{polynomial} is unbounded, but a properly defined sensitivity function should not take negative values. To introduce a lower bound at zero, we apply the soft-plus function to the sensitivities provided by the De Lange curve, $S_{DL}$:
\begin{equation}
    S_{SP}(f_t) = \ln \left[ 1 + \exp \left( S_{DL} (f_t) \right) \right].
\end{equation}
This form of parameterization provides a good fit when we use a polynomial degree of $n=3$ (goodness-of-fit: $R^2 = 0.997$, \refFig{fit_and_params_v2}).

\myfigure{fit_and_params_v2}{De Lange curve and our curve fit defined by \refEq{polynomial}. Arrows show the direction of change and the effect of an increase in the given set of curve parameters $b_i$.}

\mycfigure{model_v4}{The temporal sensitivity curves from our model for different spatial frequency contents and stimulus eccentricities on the retina.}

After fitting the curve to the measurements of De Lange, we fix the coefficients $a_i$ of this sensitivity curve for the fovea with spatially uniform luminance content (i.e., $f_h=0, f_v=0$). Then we extend this model by introducing two functions, $T(\cdot)$ and $U(\cdot)$, into the formulation for the effects of the retinal position and the spatial frequency content of the stimuli. These functions scale and shift the sensitivity curve respectively, depending on spatial frequencies ($f_h$, $f_v$) and retinal position ($e$). They are defined as
\begin{equation}
\label{eq:sensitivity}
    S(f_t, f_h, f_v, e ) = T(f_t, f_h, f_v, e) \cdot S_{SP} \Big( U \left( f_t, f_h, f_v, e \right) \Big) \cdot  ,
\end{equation}
\begin{equation}
	T( f_t, f_h, f_v, e ) = b_1 - b_2 (f_h + f_v) ^ {b_3} + b_4 e ^ {q(f_h + f_v, \bm{b}_5)} ,
\end{equation}
\begin{equation}
    U( f_t, f_h, f_v, e ) = f_t - b_6 + b_7 (f_h + f_v) + b_{8} e ,
\end{equation}
where $S ( \cdot )$ is the HVS contrast $\log$-sensitivity to a stimulus defined by spatio-temporal frequencies $f_h$, $f_v$ and $f_t$ at retinal eccentricity $e$. $B = \lbrace b_i \rbrace$ with $b_i \geq 0, \forall i \neq 5$ is the set of scalar parameters that we calibrate with psychovisual measurements. $b_2$ and $b_4$ vertically compresses the curve and $b_3$ introduces a non-linearity to the effect of spatial frequencies $f_h$ and $f_v$ on the contrast sensitivity. Another source of nonlinearity is implemented by taking into account the effect of $f_h$ and $f_v$ on the influence of retinal position $e$. This non-linearity is defined as a quadratic function to provide enough flexibility for modeling a potential non-monotonic effect:
\begin{equation}
	q(f_h+f_v, \bm{b}_5) = b_{51} (f_h+f_v)^2 + b_{52} (f_h+f_v) + b_{53},
\end{equation}
where $\bm{b}_5 = \irow{b_{51} & b_{52} & b_{53} } $ is a vector of 3 scalar parameters for the quadratic function $q(\cdot)$.

On the other hand, $U (\cdot)$ models the horizontal offset of the sensitivity curve, with $b_7$ and $b_{8}$ controlling the shift of sensitivity towards lower temporal frequencies as $f_h$, $f_v$ and $e$ increase. $b_1$ and $b_6$ adjust the position and vertical scale of the curve at the fovea, independently of the values that $f_h$, $f_v$ and $e$ take. The individual effects of the parameters $b_i$ on the behavior of the temporal sensitivity curve are shown in \refFig{fit_and_params_v2}.

\mysubsection{Decomposition of stimuli}{stimuli_decomposition}
In order to compute the visibility of visual stimuli we use Discrete Cosine Transform as our spatio-temporal frequency band decomposition method \cite{ahmed1974}. Our implementation is based on the extension of DCT-I without ortho-normalization to multiple dimensions, which is defined for one-dimensional inputs as:
\begin{equation}
\label{eq:dct}
    y_{k}=x_{0}+(-1)^{k} x_{N-1}+2 \sum_{n=1}^{N-2} x_{n} \cos \left( \frac{ \pi k n}{N-1} \right).
\end{equation}

Our method uses local DCT for windows of size $(h, w, t)$ where $h$ is the height and $w$ is the width while $t$ is the temporal length of the window. In order to compute the luminance difference $\Delta L$, DCT coefficients are scaled with a factor of $2$ in each dimension except for the coefficients with the index $k \in \lbrace 0, N-1 \rbrace$ and multiplied by the peak luminance of the display. Then the spatio-temporal band-limited Weber contrast is computed as:
\begin{equation}
	C (f_t, f_h, f_v) = \frac{\Delta L (f_t, f_h, f_v)}{\max\lbrace L (0, 0, 0), L_\text{min} \rbrace},
\end{equation}
where $f_t$, $f_h$, and $f_v$ are temporal and spatial frequency values in horizontal and vertical directions, respectively. The background luminance $L (0, 0, 0)$ is computed from the DC component of the DCT decomposition. To incorporate the effects of lower luminance levels on the contrast threshold (commonly referred to as \emph{linear} and \emph{de Vries-Rose laws} \cite{watson1986}), we clip the value of the denominator at $ L_\text{min} $.

\mysubsection{Temporal change detection probability}{temporal_change_detection}
In the studies of visual perception, the sensitivity curves represent the reciprocal of visibility threshold, where the stimuli are ``barely'' visible. This level (also known as the Just-Noticable-Difference - JND) is formally defined as 75\% probability of correctly identifying the stimulus from 2 alternatives in a psychophysical experiment, where choosing the correct alternative by random guessing is 50\%. In order to compute the probability of detection for a spatio-temporal window of visual stimulus, we first divide the contrasts computed from the DCT coefficients by the visibility thresholds given by \refEq{sensitivity}. This scales the contrasts such that a unit value corresponds to a contrast level of 1 JND:
\begin{equation}
    C_{\text{JND}} (f_t, f_h, f_v, e) =  S(f_t, f_h, f_v, e) \cdot C (f_t, f_h, f_v), 
\end{equation}
where $C (\cdot)$ is the band-limited spatio-temporal contrast computed from multidimensional DCT coefficients and $ C_{\text{JND}} (\cdot)$ is the JND-scaled contrast.

In order to compute an overall JND-scaled contrast by taking into the effect of subthreshold components of the visual stimulus, we perform spatio-temporal pooling on JND-scaled DCT contrast using Minkowski summation \cite{to2011}. We leave the DC component at $f_t = 0$ (the temporally static component of the stimulus) out of pooling:
\begin{equation}
\label{eq:pooling}
    C_{\text{M}} (e) = \left(\sum_{f_t > 0, f_h, f_v}\left| C_{\text{JND}} (f_t, f_h, f_v, e) \right|^{r}\right)^{1 / r}.
\end{equation}

Finally, we compute the probability of detecting the temporal change by applying the Weibull psychometric function \cite{weibull1951}:
\begin{equation}
    P \left( \text{detection} | C_{\text{M}} (e) \right) = p_g + \frac{(p_g-1) \cdot (1-p_l)}{\exp{\left[- \left( C_{\text{M}} (e) / \beta_0 \right) ^ {\beta_1} - 1\right]}},
\end{equation}
where $P \left( \text{detection} | C_{\text{M}} (e) \right)$ is the probability of choosing the correct alternative in a psychometric process by detecting the temporal change in a visual stimulus, $p_g$ is the guessing rate ($0.5$ in 2AFC), $p_l$ is the lapse rate (giving an incorrect answer although the stimulus is detected), $\beta_0$ and $\beta_1$ are the parameters that control the stimulus level at JND and the slope of the psychometric function, respectively.

\mysubsection{Calibration}{calibration}
\sloppy %
We calibrate the parameters of our model using the data that we collected during the perceptual experiments. The first experiment that we conducted in \refSec{experiments}, provides the thresholds measured at the selected set of spatial frequencies $f_h, f_v \in \lbrace 0\cpd, 4.5\cpd, 9.0\cpd \rbrace$ and temporal frequencies $f_z \in \lbrace 2.5 \text{Hz}, 5 \text{Hz}, 10 \text{Hz}, 20 \text{Hz}, 30 \text{Hz}, 60 \text{Hz} \rbrace $. These
thresholds are averaged among participants and used to calibrate the parameters $B = \lbrace b_i \rbrace $. The stimuli used in this experiment were synthetic sinusoidal patterns, which did not include the combined effects of multiple DCT coefficients for calibrating the pooling parameter $r$ in \refEq{pooling}. Therefore, we manually selected $r = 1.7$ for the calibration of these initial set of parameters. The temporal sensitivity curves we obtained from this first step of calibration are shown in \refFig{model_v4}.

Next, we calibrated the pooling parameter $r$ and the parameters of the psychometric function ($p_g$, $p_l$, $\beta_0$, $\beta_1$) that map the JND-scaled contrast $C_{\text{JND}}$ to the detection probability $p_d$. For this second phase of calibration, instead of synthetic stimuli, we cropped natural videos of size 71\,\si{px} $\times$ 71\,\si{px} which also include the combined effects of having different spatial and temporal frequencies, as typically observed in natural visual stimuli. In the temporal dimension, these videos consisted of 7 frames which are played back and forth in a constant loop during the experiment. We selected 3 cropped video segments for each level of JND-scaled contrast levels in $C_{\text{JND}} \in \lbrace 0.25, 0.5, 1.0, 2.0, 4.0 \rbrace$. We generated a static version of these video segments by removing all temporal frequencies except for the DC component and asked the participants to select the video with temporal changes in a 2AFC experiment where both the original and static versions are shown to the participants on the left and right parts of the display, respectively. The same participants from the previous experiment have participated in this experiment and they performed 10 repetitions for each stimulus. Next, we computed the detection rates from their responses and estimated the pooling parameter $r$ as well as psychometric function parameters ($p_g$, $p_l$, $\beta_0$, $\beta_1$) using maximum likelihood estimation. The detection rates computed from this experiment and the estimation are shown in \refFig{calibration_v3}. We tested the values for $ L_\text{min} $ from the range $ [0, 100] \;\text{cd/m}^2 $ for the fit and selected $ L_\text{min} = 50$. We provide the optimal parameter values obtained from the calibration in \refTbl{params} and the contrast threshold predictions of our model for different spatio-temporal frequencies and eccentricities are shown in \refFig{prediction_surface}. In addition, we are planning to make a Python implementation of our method publicly available for other researchers' use.

\begin{table}[hbt]
  \caption{The values of the parameters used in our model after calibration. $R^2$ is the coefficient of determination and $R^2_\text{adj}$ is the degree-of-freedom adjusted $R^2$ (number of model parameters $k=19$).}
  \label{tbl:params}
  \begin{tabular}{cccc}
  	\hline
    $a_0$ & $a_1$ & $a_2$ & $a_3$  \\
    \hline
    3.2714 & 0.3830 & 0.7669 & -0.2555  \\
    \hline
  \end{tabular}
  
  \begin{tabular}{cccc}
  	\hline
    $b_1$ & $b_2$  & $b_3$ & $b_4$  \\
    \hline
    1.0051 & 0.1830 & 0.9517 & 0.0173
    \\
    \hline
  \end{tabular}
  
  \begin{tabular}{ccccc}
  	\hline
    $\bm{b}_5$ & $b_6$ & $b_7$  & $b_8$ & r  \\
    \hline
    $\irow{-0.1375 & 0.3753 & 2.3855} $ & 0.0 & 0.0 & 0.0 & 1.9932  \\
    \hline
  \end{tabular}
  
  \begin{tabular}{cccccc}
  	\hline
    $p_g$ & $p_l$  & $\beta_0$ & $\beta_1$ & $R^2$ & $R^2_\text{adj}$  \\
    \hline
    0.5 & 0.0 & 1.7934 & 1.5 & 0.837 & 0.713  \\
    \hline
  \end{tabular}
\end{table}

\myfigure{calibration_v3}{The detection rates computed from our psychophysical experiment with the original and static natural video segments. The line shows the prediction of the detection rate from our model after calibration.}

\myfigure{prediction_surface}{The predictions of our model for spatio-temporal contrast thresholds at different retinal eccentricities.} %

\mysection{Applications}{applications}
Perceptual models and visibility predictors have a wide range of applications in computer graphics and related fields. The most simple applications include visualization and evaluation of algorithms for processing and creating visual content, e.g., rendering and compression. More advanced techniques leverage perceptual models while optimizing visual content. Due to the efficiency of our model, it can be used in both scenarios.

\myfigure{temporal_change}{Visualization of temporal change detection probabilities computed using our method for a natural video. The first frame of the video, the assumed gaze location and the overlay of computed change detection probabilities are provided in (a). Time sliced images showing the changes in temporal domain for original video (b) and the probability map (c) are shown for the red scan line in (a).}

\paragraph{Implementation and visual evaluation}
Our model (\refSec{model}) can be used to visualize the visibility of temporal changes in a video, given the gaze location from an eye tracker. Since the model operates on 71$\times$71$\times$25 spatio-temporal patches, to provide the prediction for a video, we divide the video into nonoverlapping patches of this size. For each patch, the prediction can be computed and presented in the form of a heatmap visualizing the probability of detecting the temporal changes for each spatio-temporal location. We show a sample map of change detection maps for a natural video of an ocean with waves in \refFig{temporal_change}. The computed probabilities show a declining trend as the distance from the gaze location increases. This trend is mostly attributed to the behavior observed in the HVS, which is the loss of spatio-temporal sensitivity as the retinal eccentricity increases (also observable in our model fit in \refFig{model_v4}). The processing time is 5.5 mins for a 5-second 120FPS 4K video (unoptimized parallel implementation using Python 3.6, NumPy 1.19.3, SciPy 1.5.0, OpenCV-python 4.5.1.48 on 3.6-GHz 8-core Intel Core i7-9700K CPU). The largest portion of the computational cost is incurred during the computation of DCT. Below, we provide two examples of use cases of our technique.

\mysubsection{Imperceptible transitions}{imperceptible_transitions}

\mycfigure{attention_protocol}{The overview of our application for imperceptible transitions between images. The top row shows the step size, $\Delta \alpha_n$, computed using our method to keep the probability of detection by a human observer at the specified values $ p_d \in \lbrace 0.1, 0.3, 0.5, 0.7, 0.9 \rbrace $ at the retinal positions $ e \in \lbrace 0\degree, 10\degree, 20\degree, 30\degree \rbrace $ for transitioning from a dog image to a cat image shown in the middle row (at $ \alpha_0 = 0.0 $ and $ \alpha_N = 1.0 $, respectively). Bottom row shows a sample stimuli from one of the trials, where 5 patches (randomized at each trial) are getting interpolated over time, each having a different detection probability and rate of interpolation (color coded). The experiment protocol is shown on the left of the bottom row.}

\myfigure{attention_results_v2}{Subjective experiment results for validating the visibility of images generated by our method according to the input target detection probabilities ($p_d$).}

\myfigure{attention_duration_v2}{Response times measured in our subjective experiment. We observe significantly longer response times in the experiment conducted using only the set of low temporal change detection probabilities ($L = \lbrace 0.1, 0.3, 0.5 \rbrace$, $p < 0.001$ - Wilcoxon rank-sum test). The curves represent  log-logistic probability density functions computed using MLE.}

Measuring visibility of temporal changes is important when the visibility has to be controlled within specific limits. For example, while designing graphical user interfaces for head-up or optical see-through displays, it is usually important to keep critical visual status updates more visible, whereas less critical updates should not interfere with the users' task performance by grabbing their attention unnecessarily. Similar to visible difference predictors that are designed to improve perceived quality by keeping image distortions within specific visibility limits, outputs of our method may be used for improving visual task performance and promoting sustained visual attention by adjusting the temporal visibility based on importance.

For this application, we consider a task of introducing new content into an existing scene without causing distraction to a viewer. 
We propose to consider this as a problem of computing the fastest image transition that remains undetectable when it is applied to an input image sequence at a given visual eccentricity. When transitioning from a source image to a target image, if the transition is performed slowly, the probability of detecting the visual change decreases. But using slow transitions limits how often visual information can be updated in the aforementioned applications for user interfaces or AR/VR headsets. It is possible to aim for a fast transition speed to complete the visual update in a short time, but that increases the probability of detecting the changes. A naive approach would be using a constant rate of transition not to exceed a desired probability of detection but that also requires a model to compute the probability for different transition speeds. We can perform such visual updates faster with our method because we can compute the transition speed between source and target stimuli adaptively depending on underlying content. Moreover, we can keep the probability of change detection constant over the course of the transition, making it perceptually stable.

Our method takes as input a source image ($I_s$), a target image ($I_t$), and a blending function $\phi(I_s, I_t, \alpha)$, which for $\alpha\in[0,1]$ provides a continuous transition between the two input images. Additionally, the input includes a user-chosen level of temporal change detection probability ($p_d$) and an eccentricity ($e$) at which the transition should occur. Based on the input, the method computes $\lbrace \alpha_i \rbrace_{i=1}^{N}$, such that a viewer detects the sequence of images $\lbrace I_n  = \phi(I_s, I_t, \alpha_i)\rbrace_{i=1}^{N}$ shown at the eccentricity $e$ with the probability $p_d$. The tasks is accomplished by computing the amount of increments $ \Delta \alpha_n = \alpha_{n} - \alpha_{n-1} $ that satisfies the level of detection probability, $ P_n(\text{detection} | \Delta \alpha_n) = p_d, \forall n$ at each frame update. 

To solve this problem, we use greedy optimization. We start with $\alpha_0 = 0.0 $ and compute the step sizes $ \Delta \alpha_n $ that we should take to increment $\alpha_n$ at each frame to satisfy $ P_n(\text{detection} | \Delta \alpha_n) = p_d $. In order to compute $ \Delta \alpha_n $, we apply our method to non-overlapping temporal windows of 25 video frames generated using the image blending $\phi$ and solve for the following minimization:
\begin{equation}
\label{eq:optim}
	\Delta \alpha^*_n = \argmin_{\Delta \alpha_n} \lVert P_n(\text{detection} | \Delta \alpha_n ) - p_d \rVert^2_2,
\end{equation}
where $P_n(\text{detection} | \Delta \alpha_n)$ is computed using our visibility model. To solve the above optimization problem, we apply Brent's root-finding algorithm.

Our model was calibrated using a spatial window size of $71\times71$ pixels. To compute the probability for larger image patches, we split them into smaller non-overlapping subwindows of size $71\times71$, and solve the optimization (\refEq{optim}) for each of them separately. We then apply the max-pooling strategy, which assumes that the visibility of the temporal changes in the bigger window is determined by the sub-window with the most visible changes. Consequently, we set the $\alpha_n$ to the minimum across the sub-windows.

\refFig{attention_protocol} demonstrates an example of running our optimization on a pair of cat and dog images for retinal eccentricities $ e \in \lbrace 0, 10, 20, 30 \rbrace $ and detection probabilities $ p_d \in \lbrace 0.1, 0.3, 0.5, 0.7, 0.9 \rbrace $. As the blending function here and in other our experiments, we used a linear blending $\phi(I_s, I_t, \alpha_i) = (1-\alpha_i) I_s + \alpha_i I_t$. The plots at the top of the figure visualize how the step sizes ($\Delta \alpha_n$) change depending on eccentricity and target probability. From these plots, we observe that more rapid interpolations between two images result in a higher probability of temporal change visibility. In addition, the interpolation speed defined by $\Delta \alpha_n$ is not usually uniform over time, and we see slow-downs or speed-ups depending on the image content.

The sequence of steps ($\Delta \alpha_n$) is valid only for one eccentricity value. In practice, the viewer is most likely constantly changing their gaze location, and the step sequence has to adapt to the current eccentricity value to maintain the constant level of the transition visibility. To this end, our method precomputes and stores the set of sequences ($\Delta \alpha_n$) for a finite set of different eccentricities (\refFig{attention_protocol}, top) and by smoothly interpolating between them use the sequence which corresponds to the current eccentricity. This enables a dynamic adaptation to the current gaze location. The transition slows down when the viewer's gaze is closer to the position at which the transition occurs, and conversely it speeds up when the gaze moves away. Please see our supplemental materials for experiencing the effect.

In order to evaluate our technique, we conducted a subjective experiment in which we analyzed how the optimized content impacts the participants' gaze patterns. More specifically, we were interested in validating a relation between the optimized probability of detection and eye movements towards the changing patterns. 
In each trial of the experiment, participants were shown a full-screen image containing a grid of cats and dogs images (\refFig{attention_protocol}). After a brief delay, five random patches started alternating between a cat and a dog image according to previously optimized probabilities. Participants were asked to look at the region of the image that draws their attention due to temporal changes (please see the experiment protocol in \refFig{attention_protocol}). Each trial finished as soon as the participant's gaze reached the position of one of the five changing patches. During the trial, the participant could freely move their gaze as the method was adapting the transitions according to the current gaze location.
Twelve participants (ages between 21-32) took part in the experiment conducted on an Acer X27 display at $ 3840 \times 2160 $ resolution and 120Hz refresh rate using Tobii Pro Spectrum eye tracker to monitor the gaze location. The experiment consisted of 30 trials for each participant and took approximately 5 minutes to complete.

We run two versions of this experiment. In the first version, we picked the detection probabilities of 5 pairs of patches uniformly as $ A = \lbrace 0.1, 0.3, 0.5, 0.7, 0.9 \rbrace $ (All probabilities). In the second one, we used a subset of lower probabilities $ L = \lbrace 0.1, 0.3, 0.5 \rbrace $ (Low probabilities) while keeping the number of the simultaneously changing patches during each trial the same (5). \refFig{attention_results_v2} contains a histogram of change detection probabilities ($p_d$) vs. the number of trials in which they have attracted the gaze of the participants. 

In the experiment that we tested with $ p_d \in A $, we observe that the temporal changes with $ p_d = 0.9 $ were chosen the most frequently by the participants, while this number declines rapidly as $ p_d $ decreases (\refFig{attention_duration_v2} - pink bars). We see a similar trend in the experiment with the set of $ p_d \in L $, where the participants similarly shift their gaze to the temporal change with the highest probability of detection in the set $L$ ($p_d = 0.5$) (\refFig{attention_duration_v2} - blue bars). These results demonstrate that, indeed, the higher the probability predicted by our method, the more likely the patch will attract the participant's gaze.

To further investigate the difference between the experiment with low and high probabilities, (\refFig{attention_duration_v2}) provides the time passed from the start of each trial until the participant's gaze shifts to one of the patches with temporal changes. The medians of the times are different in two experiments ($ p < 0.001 $ - Wilcoxon rank-sum test). The average time that we measured in the experiment with all probabilities is $\mu_A = 1.8951 s$ ($ CI_{95\%}: [1.5677, 2.2793]$) while the average time from the low probabilities is $\mu_L = 7.1956 s$ ($ CI_{95\%}: [6.4563, 8.2947]$) (\refFig{attention_duration_v2}). This observation suggests that although the participants shift their gaze to the patch with the highest $p_d$ shown in both experiments, there is a significant increase in the average response time, possibly due to a higher level of cognitive effort required to detect the temporal change when $p_d$ is small. We postulate that the shorter time for the experiment with all probabilities results from the fact that there were clearly visible transitions that were immediately visible to the subjects. While in the second experiment, the visibility levels were much closer to the threshold, and the participants needed more time to localize these transitions. Consequently, besides showing the effectiveness of our optimization method, this experiment further validates our model for predicting the detection probability of temporal changes in the periphery.

\mysubsection{Temporal aliasing in foveated rendering}{temporal_aliasing}

\myfigure{TAA_protocol}{The experiment protocol that we used to measure the correlation between the computed visibility of temporal changes from our method and preferences of participants (\refSec{temporal_aliasing}).}

\mycfigure{TAA_experiment_v2}{Application of our method to evaluate the temporal stability of anti-aliasing methods applied to foveated rendering. The top row shows the histograms of the computed probability of change detection from our method for the Bistro and Sponza models on the left. On the right, the average of computed probabilities is plotted against the amount of flicker perceived by the participants in our subjective experiments. The bottom row shows time-sliced images of probability maps from our method for each anti-aliasing method for the Bistro scene.}

An exciting application of our model is foveated rendering, which aims to reduce the shading rate, resolution, and bit depth to improve the rendering times or for image/video compression with minimal sacrifice of perceived quality \cite{browder1988, glenn1994, tsumura1996, kortum1996, daly2001, guenter2012}. We focus on foveated rendering applications with a lower shading rate in the periphery, which may lead to temporal aliasing. Temporal aliasing leads to deterioration in visual quality if not properly treated \cite{patney2016}. While some work has already considered modeling visibility of the foveation in static images \cite{tursun2019}, there is no technique capable of predicting the visibility of the temporal artifacts. Our method is in particular suitable for such applications. If applied directly to foveated rendering content, it can already predict visible temporal changes.

\begin{table}
	\centering
	\caption{The size of the eccentricity regions and the pixel distances that we used as a factor of native display resolution ($ 3840 \times 2160 $) in our foveated rendering implementation (\refSec{temporal_aliasing}).}
	\label{tbl:fov_regions}
	\begin{tabular}{ccc}
		\hline
		\textbf{Region} & \textbf{Radius} & \textbf{Pixel distance} \\
		\hline
		Fovea & $ 8\degree $ & $ 1 \times $ \\
		Near periphery & $ 23\degree $ & $ 2.5 \times $ \\
		Far periphery & $ 43\degree $ & $ 5.0 \times $ \\
		\hline
	\end{tabular}
\end{table}

In our experiment, we implemented our own foveated rendering testbed in the Unity game engine (HDRP - 2020.3.11f1) \shortcite{unity} with 3 eccentricity regions (i.e., fovea, near periphery, and far periphery) similar to Guenter et al. \shortcite{guenter2012} (\refTbl{fov_regions}). Then we rendered 5-second long videos of Amazon Bistro \cite{amazon_bistro} and Crytek Sponza \cite{McGuire2017Data} models with a slow camera motion in the forward direction. In different rendering runs, we applied the following anti-aliasing methods in Unity to near- and far-peripheral regions:
\begin{enumerate}
	\item Fast approximate anti-aliasing (FXAA) \cite{lottes2009} 
	\item Subpixel morphological anti-aliasing (SMAA) \cite{jimenez2012} (quality preset: high)
	\item Temporal anti-aliasing (TAA) \cite{korein1983} (quality preset: high)
\end{enumerate}
In addition to these anti-aliasing methods, we also rendered both models without applying any anti-aliasing (No AA) and computed the probability of temporal change detection from all videos using our method.

In order to measure the correlation of probabilities computed by our method and the visibility of any temporal artifacts in the output of anti-aliasing methods, we conducted a 2AFC subjective experiment, where the participants compared pairs of videos that we rendered. The same group of participants that has participated in the experiment of imperceptible transitions (\refSec{imperceptible_transitions}) did this experiment. It consisted of 12 trials, and in each trial the participants were asked to watch a pair of anti-aliasing results from the same scene and choose the one with less flickering (\refFig{TAA_protocol}). The experiment was conducted on the same 55-inch LG OLED55CX, 120\si{Hz}, 4K display that we used to calibrate our model due to its large field-of-view (\refSec{experiments}). The pairwise comparison results from this subjective experiment was converted into just-objectionable-difference (JOD) quality scores using Thurstonian scaling \cite{thurstone1927, perez2017}. The probability maps of temporal change detection from our method are pooled using Minkowski summation with exponent $\beta = 3$ to obtain a scalar score \cite{graham1978, rohaly1997, to2011}. The histogram of the probabilities computed for each anti-aliasing method and a plot of the JOD scores computed from the subjective experiment vs. pooled probabilities from our method are shown in \refFig{TAA_experiment_v2}. We observe that FXAA and SMAA methods scored close to the rendering result with no anti-aliasing, whereas TAA turned out to be significantly superior for suppressing flickering in the periphery according to the subjective experiment results. The average probability of temporal change detection computed by our method is also in agreement with the results of subjective experiment (Pearson $\rho = -0.903$, $ p = 0.002$ - t-test). Upon visual inspection, we also observe that the computed probability maps overall show higher probability of change detection for No AA, FXAA, and SMAA compared to TAA (please see the time-sliced images at the bottom row of \refFig{TAA_experiment_v2}).

A direct application of our method to natural videos would detect the temporal changes that also arise from motion in the scene. Under some circumstances, it may be desirable to evaluate the potential aliasing due to only foveation. We also show an application that decouples the temporal changes due to motion in the scene and the aliasing. To this end, we warp the subsequent frames using motion flow vectors, effectively removing any motion, before applying our model (\refFig{aliasing_v2}). As it can be observed in the figure, when such compensation is not performed, the visibility of the aliasing is dominated by the motion. Both with and without anti-aliasing sequences produce similar visibility maps (top row). When the motion compensation is applied, only the effect of aliasing is detected by our method. Consequently, the prediction for the sequence with motion compensation and anti-aliasing does not include visible temporal changes.

\myfigure{aliasing_v2}{Application of our technique to the analysis of temporal aliasing with and without motion compensation. Dashed circles are the boundaries of foveal, near-peripheral and far-peripheral regions. The pixel distance used in each region is shown on the images as a factor of native display resolution.}

\mysection{Discussion}{Discussion}

\myfigure{comparison_krajancich}{Comparison of critical flickering frequency (CFF) computed by our method and Krajancich et al.\shortcite{Krajancich2021}. Straight lines correspond to the part of the curves obtained by fitting to actual measurements, whereas dashed lines represent extrapolations of the models. The bounds of the measured eccentricities are computed as the summation of reported eccentricity and the Gaussian window parameter ($\sigma$) for Krajancich et al.\shortcite{Krajancich2021} because their stimuli size depends on the spatial frequency level tested.}

\myfigure{fovvdp_comparison}{The probability of detecting a visible temporal change by a human observer as estimated by FovVDP and our method. The histograms on the left show the predictions from two methods for cross-modulating sinusoidal gratings (\refSec{experiments}), whereas the scatter plots on the right show the predictions for complex stimuli (\refSec{calibration}). $ \beta = 3$ is the Minkowski summation parameter used for global pooling of our method's predictions (for details, please refer to the text).}

We provide a discussion comparing our model with two recent works on similar topics. In the first one, Krajancich~et~al.~\shortcite{Krajancich2021} provide measurements and a model of critical flicker frequencies across a wide visual field of view (up to 60$^{\circ}$ of eccentricity) for Gabor patches of spatial frequencies up to 2\cpd. To handle higher spatial frequencies, the model relies on extrapolation using an existing model for spatial acuity. Compared to their work, our measurements aim at acquiring sensitivity of the HVS to continuous spatio-temporal signal variation. While our measurements cover a slightly lower range of eccentricities (45$^{\circ}$), we test spatial frequencies up to 15\cpd. More importantly, in contrast to the suggested application of Krajancich~et~al. based on Discrete Wavelet Transform (DWT), we show end-to-end applications with DCT-based video decomposition and its thorough subjective validation in two experiments. DCT has no special advantage over other well-known band decomposition methods using Fourier or Gabor basis functions because complex stimuli can be represented equally well in all three approaches. We opted for DCT decomposition in our technique because it has established widespread use in image compression standards such as JPEG and there is a support from a large variety of numerical libraries. That provides convenience while implementing complex video content processing tasks, some of which are demonstrated in this paper.

We compute CFF with our method and compare it with the data provided by Krajancich~et~al. in \refFig{comparison_krajancich}. Both models are calibrated to combinations of different retinal positions of stimuli (eccentricity) and spatial frequency content that are presented during psychovisual experiments (\refFig{comparison_krajancich}, solid lines). Outside the region of measurements, there is no experimental data for fitting the model parameters and CFF computations are obtained by extrapolation (\refFig{comparison_krajancich}, dashed lines). One major difference between our study and Krajancich~et~al. is the stimuli size, which is fixed in our experiments whereas it grows to keep the number of Gabor cycles constant in Krajancich~et~al. The difference becomes significant especially for low spatial frequencies, because the stimuli cover the whole visual field when the spatial frequency content reaches zero in the experiments of Krajancich~et~al. Such stimuli essentially test all retinal eccentricities at the same time. In contrast, we always use a fixed envelope for sinusoidal gratings that make our measurements local. Another difference between two studies is the type of displays used in the experiments. While the study of Krajancich~et~al. is able to measure temporal frequencies up to and beyond 100\Hz, we made our measurements on a display that supports up to 60\Hz (120FPS). Our model is in agreement with the predictions of Krajancich~et~al. for Krajancich~et~al. predict an increase in CFF for 2\cpd. Beyond this frequency, their model relies on extrapolation. On the other hand, for spatial frequencies below 2\cpd, their model predicts an increase in CFF with eccentricity before it declines again in the far periphery. This is an observation commonly shared by previous CFF studies. In contrast, we observe that our model predicts a monotonic decrease in CFF as the stimulus eccentricity increases even when the spatial frequency is below 2\cpd. This type of behavior in our model may be explained by a fixed and relatively small stimuli size that results in a monotonic change in CFF with eccentricity \cite{hartmann1979}.

In the second work, a different problem is addressed by Mantiuk~et al.~\shortcite{Mantiuk2021}. The authors propose a quality metric (FovVDP) for wide field-of-view videos. The method is trained on a dataset containing information about comfort and uniformity of quality degradation obtained using an off-the-shelf virtual reality headset. Compared to our technique, their method targets different applications. It aims to predict the overall quality score and supra-threshold visibility of a change in the quality with respect to a reference video by providing a scalar quality score for the entire sequence. In contrast, the goal of our model is a precise prediction and localization of the probability of seeing local temporal changes without a reference. This is critical for many applications in computer graphics where localization is important and having a non-reference method is desirable. While the method by Mantiuk~et~al. provides error distribution across each frame, the interpretation of these values remains difficult, as the metric was not trained on a local visibility dataset. Another important aspect of our technique is that our precise visibility measurement will more easily extend across different display devices than the calibration based on a dataset collected on a rather limited headset.

We compare the outputs from our method and from Mantiuk~et~al. on our datasets of cross-modulating sinusoidal gratings (\refSec{experiments}) and complex stimuli (\refSec{calibration}). We select our dataset as the common input for both methods because there is no established dataset from previous works for performing such a comparison. We convert the scalar Just Objectionable Difference (JOD) score produced by FovVDP to the probability of detection using the inverse of the standard normal CDF. In contrast to FovVDP, our method does not have a global pooling step, and it is calibrated to predict local visibility across the visual field. Therefore, we apply Minkowski summation with parameter $ \beta = 3 $ \footnote{for an actual application, $ \beta $ should be estimated based on new psychovisual experiment data, which is left out of the scope of this study}, which is a reasonable value for visual cue summation \cite{to2011}. The reference input of FovVDP consists of a uniform gray level that is equal to the background (also the temporal average of the stimuli). The modulation amplitude of the stimuli is set to the threshold from our experiment that corresponds to the detection probability $ P(\text{det}) = 0.5 $. We show the histogram of the probability predictions from FovVDP and our method in \refFig{fovvdp_comparison}. We observe that FovVDP predictions significantly underestimate visibility with a peak close to $ P(\text{det}) = 0 $. On the other hand, the histogram of the predictions from our method resembles a truncated Gaussian with a mean around $ P(\text{det}) = 0.2 $. This is still an underestimation because, ideally, both methods should produce predictions centered around $ P(\text{det}) = 0.5 $. For our method, we attribute this observation to how visual masking is handled. Our method does not explicitly model visual masking effects, but it is calibrated on complex stimuli that include masking effects. We believe that this leads to underestimated predictions for simple stimuli that have a very sparse DCT representation. On the right side of \refFig{fovvdp_comparison}, we show the predictions of FovVDP and our method for complex stimuli from \refSec{calibration}. We observe that FovVDP predicts most of the stimuli as barely visible unless $ P(\text{det})$ is very close to 1. Our predictions show a smoother transition as the measured probability increases.

To summarize, the two mentioned techniques and our method aim to model human perception in peripheral vision. Yet, the differences between them, make them suitable for different applications. While work by Mantiuk~et~al. focuses on quality score for an entire video sequence, our method can provide a precise information about the local visibility of temporal differences. In addition, different from our method, their predictor is not trained on a dataset of temporal change visibility. On the other hand, perceptual model by Krajancich~et~al. addresses the problem of critical flicker frequencies, which is more similar to our goal, but it is unclear at this point how it can be applied to complex video content.

\mysection{Limitations and future work}{future_work}

From vision science perspective, peripheral stimuli are scaled according to the cortical magnification factor (CMF) and the envelope of spatial gratings is selected such that it contains at least 2--3 cycles \cite{howell1978,virsu1982,watson87,johnston87}. On the other hand, we used a fixed stimuli size in our experiments because of our applications, which use DCT decomposition with a constant window size. The influence of limiting the envelope size for stimuli with low spatial frequencies is not clearly modeled by previous studies. In future work, this effect may be investigated with a frequency band decomposition that allows for variable envelope sizes.

In our experiments, we did not consider different backgrounds. As a result, we did not model the spatial visual masking effect. Masking can potentially reduce visibility; therefore, without this consideration, our model remains conservative in its prediction. Investigating the effect of masking in the experiments and collecting samples for a wider range of temporal and spatial frequencies, as well as different luminance levels, will lead to a more accurate model. However, when designing such experiments, it is critical to monitor the dimensionality of the problem because as more factors are investigated, psychophysical experiments may quickly become infeasible in terms of duration.

Another limitation regarding our experiments is the number of participants that did not allow us to capture the variability of the measured thresholds in the population and increase the noise in our measurement. However, we argue that because of the nature of our experiments, the important perceptual characteristic is captured in the measurements. Furthermore, the utilization of previous measurements for fovea \cite{de_lange1952} allowed us to regularize our possibly noisy data. Having a small number (4) of participants to model and calibrate perceptual models is not uncommon in academic vision science. This is partly due to long experiment sessions (12 $\times$ 20 minutes) and more extensive measurements performed in a controlled experiment environment (e.g., display and ambient luminance levels, display calibration, proper positioning of the viewer, avoiding participant fatigue, vision tests, etc.). On the other hand, for industrial applied vision, general practice aims for a higher number of participants when feasible. In our measurements, the data collected from different participants were mostly in agreement with each other, and we did not have conflicting observations in the data that would otherwise require seeking additional participants or revising the experiment protocol. Moreover, the calibrated perceptual model is verified with further experiments described in applications section with a larger participant group, which would otherwise fail if the model was not representative of general perceptual characteristics of the HVS.

When modeling our data from psychophysical experiments, we rely on previous measurements derived for fovea \cite{de_lange1952}. While we observe a good fit of our data, our model extrapolates beyond our measurements. Therefore, the accuracy could be improved with more extensive measurements. Furthermore, the choice of temporal CSF data could be improved with those newer than De~Lange~\shortcite{de_lange1952} such as the data provided by Watson~\shortcite{watson1986}.

Our model applies pooling only across different DCT components. It was a design choice not to include spatial pooling across different patches since we did not collect the necessary data. In future work, when threshold measurements for different sizes of stimuli are performed, an improved version of our model could use a multi-scale approach to account for temporal fluctuations with different spatial support.

Imperceptible transitions, which is discussed as one of our applications in \refSec{imperceptible_transitions}, have been extensively studied in the past as a part of video compression and watermarking \cite{bi2013,lubin2003,bradley2012,noorkami2008}. However, most of those studies do not model visibility as a function of eccentricity. An application of our model to concealing digital watermarks in videos and adjusting video compression rates based on temporal change visibility may be promising future research directions.

Finally, each one of our applications is a demonstration of a possible use case of our model. In our work, we focused on deriving the model while leaving the full development of applications that utilize it for future work. A potential research direction would be towards decoupling temporal changes (e.g., due to motion and other factors), then computing their individual contributions to the visibility. Another interesting research direction would be to control the speed of complex image interpolation techniques which consist of multiple transformations to the inputs (e.g., a combination of image warping and interpolation). Such methods would require an extension of our application on imperceptible image transitions (\refSec{imperceptible_transitions}) to an optimization of multi-dimensional $\Delta \alpha$ vector.

\mysection{Conclusion}{conclusion}
With the development of new wide-field-of-view displays equipped with eye tracking technology, correct treatment of peripheral vision becomes essential. In this work, we argue that perceptual models for the peripheral vision that address spatial and temporal aspects of our perception will lead to more efficient image generation techniques and new applications that will contribute to the final user experience. With this goal in mind, we presented experiments that investigate the visibility of temporal changes in the periphery. The stimuli are chosen with the goal of incorporating multiple characteristics such as temporal and spatial frequencies as well as eccentricity.
These characteristics are essential to enable the modeling of the perception of complex content. Using our measurements, we proposed a novel model that can predict the visibility of temporal fluctuations in complex content. We also discussed and presented examples of possible applications. While the current model can be already successfully applied to various computer graphics applications, we hope that this work will lead to further developments of more comprehensive models which address both spatial and temporal aspect of our perception across a wide field of view.

\begin{acks}
This project has received funding from the European Research Council (ERC) under the European Union’s Horizon 2020 research and innovation program (grant agreement N◦ 804226 PERDY).
\end{acks}

\appendix

\section{The power transformation}
\label{appendix:power_transform}

In order to avoid the mathematical singularity at $\log(0)$, instead of using the standard $\log$-transformation, we use the two-parameter Box-Cox power transformation on spatial and temporal frequencies $f$:
\begin{equation}
	\tilde{f}=\left\{\begin{array}{ll}
		\ln \left(f+\lambda_{2}\right) & \text { if } \lambda_{1}=0, \\
		\frac{\left(f+\lambda_{2}\right)^{\lambda_{1}}-1}{\lambda_{1}} & \text { if } \lambda_{1} \neq 0,
	\end{array}\right.
\end{equation}
where $\tilde{f}$ is the transformed frequency $f$. It is also applied to spatio-temporal sensitivities $S$ instead of the standard $\log$ transformation, where it is required. This transformation has a nice property of keeping zeroes intact in the $\log$ domain with $\lambda_1 = 0$ and $\lambda_2 = 1$.

\section{Cross Validation of Function Fit
\label{appendix:cross_validation}}
We provide 5-fold cross-validation results for our calibration in \refSec{calibration} and parameters estimated for each fold in \refTbl{crossval}. We observe close train and test losses, which suggests that an overfit is unlikely. In addition, estimated parameter values appear to be stable between cross-validation folds.

\begin{table}[h]
	\centering
	\caption{Cross validation results of our calibration in \refSec{calibration}.}
	\label{tbl:crossval}
	\resizebox{1.0\columnwidth}{!}{%
		\begin{tabular}{rrrrrrrrrrrrr}
			\hline
			\textbf{CV-fold} & \textbf{$\mathcal{L}_\text{train}$} & \textbf{$\mathcal{L}_\text{test}$} & \textbf{$b_1$} & \textbf{$b_2$} & \textbf{$b_3$} & \textbf{$b_4$} & \textbf{$b_{5,1}$} & \textbf{$b_{5,2}$} & \textbf{$b_{5,3}$} & \textbf{$b_6$} & \textbf{$b_7$} & \textbf{$b_8$} \\
			\hline 
			1 & 0.122 & 0.091 & 1.008 & 0.208 & 0.892 & 0.008 & -0.146 & 0.389 & 2.952 & 0.000 & 0.030 & 0.054 \\
			2 & 0.119 & 0.126 & 1.015 & 0.140 & 1.128 & 0.025 & -0.152 & 0.497 & 1.988 & 0.000 & 0.050 & 0.055 \\
			3 & 0.119 & 0.142 & 1.015 & 0.176 & 0.991 & 0.021 & -0.145 & 0.409 & 2.241 & 0.000 & 0.033 & 0.049 \\
			4 & 0.107 & 0.146 & 1.013 & 0.123 & 1.200 & 0.033 & -0.143 & 0.452 & 1.856 & 0.000 & 0.051 & 0.055 \\
			5 & 0.102 & 0.133 & 1.015 & 0.161 & 1.039 & 0.022 & -0.146 & 0.435 & 2.179 & 0.000 & 0.020 & 0.064 \\
			\hline 
			\textbf{Mean} & 0.114 & 0.127 & 1.013 & 0.162 & 1.050 & 0.022 & -0.146 & 0.436 & 2.243 & 0.000 & 0.037 & 0.055 \\
			\textbf{Stdev} & 0.009 & 0.022 & 0.003 & 0.033 & 0.120 & 0.009 & 0.003 & 0.041 & 0.425 & 0.000 & 0.013 & 0.005
		\end{tabular}%
	}
\end{table}

\bibliographystyle{ACM-Reference-Format}
\bibliography{ms.bib}

\end{document}